\documentclass[%
 reprint,
 superscriptaddress,
 amsmath,amssymb,
 aps,
pre,
]{revtex4-2}

\usepackage{graphicx}
\usepackage{dcolumn}
\usepackage{bm}

\begin{document}

\preprint{APS/123-QED}

\title{Frequency Extraction from Invariant Flows}

\author{Derong Xu}\email{dxu@bnl.gov}
\author{Yongjun Li}%
\affiliation{Brookhaven National Laboratory}
\author{Yue Hao}%
\affiliation{Michigan State University}
\author{Sergei Nagaitsev}%
\affiliation{Brookhaven National Laboratory}

\date{\today}

\begin{abstract}
In non-degenerate integrable Hamiltonian systems, invariant tori can be parameterized equivalently by action variables or by their fundamental frequencies. We introduce an invariant-flow formulation for extracting fundamental frequencies of integrable Hamiltonian systems. By treating invariants as generators of commuting Hamiltonian flows, the frequencies are obtained from time-of-flight parameters along these flows, providing a direct alternative to action-angle constructions and spectral methods based on long time series. The approach yields an explicit numerical procedure that extends naturally to systems with multiple degrees of freedom. Its effectiveness is demonstrated using the McMillan map, where machine-precision accuracy is achieved.
\end{abstract}

\maketitle


Integrable Hamiltonian systems possess as many independent, commuting invariants as degrees of freedom, and these invariants define smooth tori in the phase space. The quasi-periodic evolution on each torus is characterized by a set of fundamental frequencies. Extracting these frequencies directly from invariants provides an alternative to action-angle constructions, which are often difficult to obtain explicitly.

In realistic accelerator applications, circular machines are designed to operate in regimes that remain close to integrability in order to preserve large dynamic aperture and long-term beam stability. The KAM theorem \cite{dumas2014kam} implies that a large fraction of invariant tori persist under small perturbations. In such near-integrable settings, exact or approximate invariants can be constructed either from data-driven approaches \cite{PhysRevLett.126.180604} or directly from the one-turn map \cite{m349-wmnr}.

This Letter presents a theoretical framework and a practical algorithm for extracting fundamental frequencies directly from the invariants of integrable systems.

The Liouville-Arnold theorem states that an integrable Hamiltonian system can be described in action-angle variables, in which the actions are constants of motion and the angles evolve as rigid rotations with fixed frequencies.
In the one-degree-of-freedom (1DOF) case, this yields an explicit expression for the frequency (rotation number) in terms of the invariant.
The Danilov equation expresses the frequency as \cite{PhysRevAccelBeams.23.054001}
\begin{equation}
    \nu\left(\mathcal{K}\right)=\left.{\int_{q_0}^{q_1} \left(\frac{\partial \mathcal{K}}{\partial p}\right)^{-1}\mathrm{d}q}\right/{\oint \left(\frac{\partial \mathcal{K}}{\partial p}\right)^{-1}\mathrm{d}q}
    \label{eq:Danilov}
\end{equation}
Here $(q,p)$ are canonical coordinates, $(q_0,p_0)$ is an initial condition, and $(q_1,p_1)$ is its image under the one-turn map.
The integrals are taken along the invariant curve $\mathcal{K}(q,p)=\mathrm{const}$.
However, extending this formulation beyond 1DOF is not straightforward.

Equation~(\ref{eq:Danilov}) can be interpreted as a time-of-flight ratio by viewing the invariant $\mathcal{K}$ as a Hamiltonian generating a flow along its level set.
Under this flow, the numerator gives the time required to transport the phase-space point from $(q_0,p_0)$ to its image $(q_1,p_1)$, while the denominator gives the period of the closed orbit on the invariant curve $\mathcal{K}(q,p)=\mathrm{const}$.
The frequency is the ratio of these two times.

This observation motivates a flow-based construction of frequencies that generalizes beyond 1DOF.
In the following, without loss of generality, we illustrate the method using a two-degree-of-freedom (2DOF) system, since the construction depends only on the existence of commuting invariants. The framework extends directly to higher-dimensional integrable systems.

The Hamiltonian of a 2DOF integrable system can be written in action-angle variables $(\phi_1,J_1,\phi_2,J_2)$ as
\begin{equation}
    \mathcal{H} = \mathcal{H}(J_1,J_2).
\end{equation}
The corresponding fundamental frequencies are
\begin{equation}
    \nu_i = \frac{1}{2\pi}\frac{\partial \mathcal{H}}{\partial J_i},
    \qquad i=1,2 .
    \label{eq:nu}
\end{equation}
The explicit form of $\mathcal{H}$ and the associated actions are generally unavailable.
Instead, the dynamics is accessed through the discrete symplectic map generated by the Hamiltonian flow,
\begin{equation}
    \Phi_\mathcal{H}^t=\exp\!\left(-t\colon \mathcal{H} \colon\right),
    \qquad t=0,1,2,\ldots,
    \label{eq:Hflow}
\end{equation}
where $\colon \mathcal{H} \colon f=[\mathcal{H},f]$ denotes the Poisson bracket for an arbitrary function $f$.

Given two independent integrals of motion,
\begin{equation}
\mathcal{K}_1=\mathcal{K}_1(J_1,J_2),
\qquad
\mathcal{K}_2=\mathcal{K}_2(J_1,J_2),
\label{eq:K1K2}
\end{equation}
their explicit dependence on the actions is generally unknown.
Instead, their expressions in terms of the phase-space coordinates $(x,p_x,y,p_y)$ are available.

By the chain rule, the Poisson bracket with the Hamiltonian can be written as
\begin{equation}
    [\mathcal{H},f]
    =
    \frac{\partial \mathcal{H}}{\partial \mathcal{K}_1}[\mathcal{K}_1,f]
    +
    \frac{\partial \mathcal{H}}{\partial \mathcal{K}_2}[\mathcal{K}_2,f],
\end{equation}
which leads to the decomposition of the operator,
\begin{equation}
    \colon \mathcal{H}\colon
    =
    \frac{\partial \mathcal{H}}{\partial \mathcal{K}_1}\colon \mathcal{K}_1 \colon
    +
    \frac{\partial \mathcal{H}}{\partial \mathcal{K}_2}\colon \mathcal{K}_2\colon .
\end{equation}

Since the flows generated by $\mathcal{K}_1$ and $\mathcal{K}_2$ commute, the Baker--Campbell--Hausdorff (BCH) formula gives the factorization of the Hamiltonian flow,
\begin{equation}
    \begin{aligned}
    \exp\left(-t\colon \mathcal{H}\colon\right)=
    \exp&\!\left[-t\!\left(\frac{\partial \mathcal{H}}{\partial \mathcal{K}_1}\right)\! \colon \mathcal{K}_1\colon\right]\\
    &\exp\!\left[-t\!\left(\frac{\partial \mathcal{H}}{\partial \mathcal{K}_2}\right)\! \colon \mathcal{K}_2\colon\right].
    \end{aligned}
\end{equation}
Defining
\begin{equation}
    \tau_1(t)=t\left(\frac{\partial \mathcal{H}}{\partial \mathcal{K}_1}\right),
    \qquad
    \tau_2(t)=t\left(\frac{\partial \mathcal{H}}{\partial \mathcal{K}_2}\right),
\end{equation}
the $\mathcal{H}$-flow can be written compactly as a composition of invariant flows,
\begin{equation}
    \Phi_\mathcal{H}^t
    =\Phi_{\mathcal{K}_1}^{\tau_1}\Phi_{\mathcal{K}_2}^{\tau_2}
    =\Phi_{\mathcal{K}_2}^{\tau_2}\Phi_{\mathcal{K}_1}^{\tau_1}.
\end{equation}

The flows generated by $\mathcal{H}$, $\mathcal{K}_1$, and $\mathcal{K}_2$ preserve the actions and induce rigid rotations but with distinct angular velocities. For the factorized representation to reproduce the same one-turn evolution ($t=1$ in Eq.~\ref{eq:Hflow}), the angle increments generated by the composition of the $\mathcal{K}_1$ and $\mathcal{K}_2$ flows must match those produced by the $\mathcal{H}$-flow. Consistency therefore requires
\begin{equation}
\renewcommand{\arraystretch}{1.5}
\begin{pmatrix}
    \partial \mathcal{K}_1/\partial J_1
    &
    \partial \mathcal{K}_2/\partial J_1
    \\
    \partial \mathcal{K}_1/\partial J_2
    &
    \partial \mathcal{K}_2/\partial J_2
\end{pmatrix}
\begin{pmatrix}
\tau_1
\\
\tau_2
\end{pmatrix}
=
\begin{pmatrix}
    \partial \mathcal{H}/\partial J_1
    \\
    \partial \mathcal{H}/\partial J_2
\end{pmatrix}.
\label{eq:phi12}
\end{equation}

The motion is confined to a smooth two-dimensional invariant torus. In the non-resonant case, the trajectory on a given torus is quasi-periodic and densely fills the torus. The flows generated by $\mathcal{H}$, $\mathcal{K}_1$, and $\mathcal{K}_2$ remain on the same torus and translate the phase-space point along directions defined by their Hamiltonian vector fields \((\dot{x},\dot{p}_x,\dot{y},\dot{p}_y)\). 
Because $\mathcal{K}_1$ and $\mathcal{K}_2$ are independent invariants that Poisson-commute, their flows generate two linearly independent directions on the torus. By composing these flows with appropriate time-of-flight parameters, any point on the same invariant torus can be reached. In particular, suitable combinations generate closed loops on the torus,
\begin{equation}
    \Phi_{\mathcal{K}_1}^{T_1}\Phi_{\mathcal{K}_2}^{T_2}
    =
    \exp\!\left[-2\pi:\left(m_1J_1+m_2J_2\right):\right],
\end{equation}
where $m_{1,2}$ are integers specifying the winding numbers associated with the actions $J_{1,2}$, and $T_{1,2}$ are the corresponding time-of-flight parameters under the $\mathcal{K}_{1,2}$ flows.
Consistency with Eq.~(\ref{eq:phi12}) then yields
\begin{equation}
\renewcommand{\arraystretch}{1.5}
\begin{pmatrix}
    \partial \mathcal{K}_1/\partial J_1
    &
    \partial \mathcal{K}_2/\partial J_1
    \\
    \partial \mathcal{K}_1/\partial J_2
    &
    \partial \mathcal{K}_2/\partial J_2
\end{pmatrix}
\begin{pmatrix}
T_1
\\
T_2
\end{pmatrix}
=
2\pi
\begin{pmatrix}
m_1
\\
m_2
\end{pmatrix}.
\end{equation}

Choosing two linearly independent integer vectors $(m_{11},m_{21})$ and $(m_{12},m_{22})$, together with the corresponding time-of-flight solutions $(T_{11},T_{21})$ and $(T_{12},T_{22})$, resolves the Jacobian matrix as
\begin{equation}
\renewcommand{\arraystretch}{1.5}
\begin{pmatrix}
    \partial \mathcal{K}_1/\partial J_1
    &
    \partial \mathcal{K}_2/\partial J_1
    \\
    \partial \mathcal{K}_1/\partial J_2
    &
    \partial \mathcal{K}_2/\partial J_2
\end{pmatrix}
=
2\pi
\begin{pmatrix}
m_{11} & m_{12}\\
m_{21} & m_{22}
\end{pmatrix}
\begin{pmatrix}
T_{11} & T_{12}\\
T_{21} & T_{22}
\end{pmatrix}^{-1}.
\label{eq:coef}
\end{equation}
Combining Eqs.~(\ref{eq:coef}), (\ref{eq:phi12}), and (\ref{eq:nu}) gives the fundamental frequencies,
\begin{equation}
\renewcommand{\arraystretch}{1.5}
\begin{pmatrix}
\nu_1\\
\nu_2
\end{pmatrix}
=
\begin{pmatrix}
m_{11} & m_{12}\\
m_{21} & m_{22}
\end{pmatrix}
\begin{pmatrix}
T_{11} & T_{12}\\
T_{21} & T_{22}
\end{pmatrix}^{-1}
\begin{pmatrix}
\tau_1\\
\tau_2
\end{pmatrix}.
\label{eq:sol}
\end{equation}

The parameters $\tau_1$ and $\tau_2$ are determined from symplectic integration of the invariant flows. Let $\mathbf{z}=(x,p_x,y,p_y)^\mathrm{T}$ denote the phase-space vector and $\mathbf{z}_0$ an initial point.
Applying the one-turn map yields the image $\Phi_{\mathcal{H}}(\mathbf{z}_0)$.
For trial values of $\tau_1$ and $\tau_2$, a reconstructed point is obtained by composing the invariant flows,
\begin{equation}
    \mathbf{z}\left(\tau_1,\tau_2\right)=\Phi_{\mathcal{K}_2}^{\tau_2}\left(\Phi_{\mathcal{K}_1}^{\tau_1}\left(\mathbf{z}_0\right)\right)=\Phi_{\mathcal{K}_1}^{\tau_1}\left(\Phi_{\mathcal{K}_2}^{\tau_2}\left(\mathbf{z}_0\right)\right)
\end{equation}
The parameters \((\tau_1,\tau_2)\) are then determined by minimizing the mismatch
between the reconstructed point and the true one-turn image,
\begin{equation}
    {L}(\tau_1,\tau_2)=\bigl\|\mathbf{z}\left(\tau_1,\tau_2\right)-\Phi_\mathcal{H}(\mathbf{z}_0)\bigr\|.
\end{equation}
The invariant flows are evaluated numerically, preserving symplectic structure, and the minimization can be performed with a gradient-based optimizer.
The derivatives $\partial \mathbf{z} / \partial \tau_i$ are available analytically,
\begin{equation}
    \frac{\partial \mathbf{z}}{\partial \tau_i}
    =
    [\mathbf{z},\mathcal{K}_i]\bigg|_{\mathbf{z}=\mathbf{z}(\tau_1,\tau_2)},
    \qquad i=1,2.
\end{equation}

Closed loops on the invariant torus are constructed analogously by seeking
time-of-flight parameters $(T_1,T_2)$ such that the composition of the invariant
flows returns the phase-space point to its initial location,
\begin{equation}
    {L}(T_1,T_2)
    =
    \bigl\|\mathbf{z}\left(T_1,T_2\right)-\mathbf{z}_0\bigr\|.
\end{equation}
Different solutions correspond to distinct closed cycles on the torus and are
obtained by varying the initial guess of the optimizer.
Once the parameters $\tau_i$ and the independent closed-loop times $T_{ij}$ are
determined, the fundamental frequencies follow directly from
Eq.~(\ref{eq:sol}).

The integer matrix $m_{ij}$ appearing in Eq.~(\ref{eq:sol}) remains to be determined.
Let $\mathbf{m}\in\mathbb{Z}^{2\times 2}$ collect two independent winding vectors, and let
$\mathbf{T}\in\mathbb{R}^{2\times 2}$ collect the corresponding time-of-flight
parameters.
Equation~(\ref{eq:sol}) can then be written compactly as
\begin{equation}
    \boldsymbol{\nu}=\mathbf{m}\,\mathbf{T}^{-1}\boldsymbol{\tau}.
    \label{eq:nuvector}
\end{equation}
The fundamental frequencies are defined with respect to a chosen basis of cycles
on the invariant torus, taken here as the elementary windings $(1,0)$ and $(0,1)$
corresponding to one turn around the actions $J_1$ and $J_2$, respectively.

The choice of cycle basis on the torus is not unique.
A change of basis is represented by an unimodular integer matrix
$\mathbf{U}\in\mathrm{GL}(2,\mathbb{Z})$, under which the winding matrix transforms as
\begin{equation}
    \mathbf{m}'=\mathbf{m}\mathbf{U}.
    \label{eq:mU}
\end{equation}
If the time-of-flight matrix is transformed consistently,
\begin{equation}
    \mathbf{T}'=\mathbf{T}\mathbf{U},
    \label{eq:TU}
\end{equation}
then
\begin{equation}
    \mathbf{m}'\,\mathbf{T}'^{-1}
    =
    \mathbf{m}\,\mathbf{T}^{-1},
    \label{eq:invariantproduct}
\end{equation}
and the frequency vector \(\boldsymbol{\nu}\) in Eq.~(\ref{eq:nuvector}) remains
the same.

In practical applications, the time-of-flight matrix $\mathbf{T}$ is obtained
independently of the choice of cycle basis, whereas the integer matrix
$\mathbf{m}$ is not known {a priori}.
Holding $\mathbf{T}$ fixed and replacing
$\mathbf{m}\rightarrow\mathbf{m}'=\mathbf{m}\mathbf{U}$ therefore produces
different candidate frequency vectors
$\boldsymbol{\nu}'=\mathbf{m}'\mathbf{T}^{-1}\boldsymbol{\tau}$.
This ambiguity arises solely from the choice of cycle basis used to represent the
motion on the invariant torus and is independent of the underlying dynamics.
It therefore motivates the calibration of $\mathbf{m}$ using additional
information.

Even restricting the frequencies to the range
$0<\nu_{1,2}<0.5$ does not uniquely determine the integer matrix $\mathbf{m}$.
To fix this ambiguity, a coarse estimate can be obtained analytically from the linear one-turn map.
Although such estimates are not sufficiently precise for high-accuracy frequency
determination, they are adequate to discriminate among the finite set of
candidate integer matrices.
Only one choice yields a frequency vector consistent with this estimate, thereby
fixing the cycle basis unambiguously.

Once the integer matrix has been calibrated for a given torus, it can be reused
for nearby tori, since the frequencies vary smoothly with the actions.
The corresponding time-of-flight parameters then provide an excellent initial
guess for the optimization used to determine the frequencies on nearby tori.

As a demonstration, we apply the method to the two-dimensional McMillan map
\cite{McMillan1971}, whose one-turn map is
\begin{equation}
    \begin{aligned}
        x^{\mathrm{new}} &= p_x, &
        p_x^{\mathrm{new}} &= -x + \frac{a p_x}{1 + b\left(p_x^2 + p_y^2\right)}, \\
        y^{\mathrm{new}} &= p_y, &
        p_y^{\mathrm{new}} &= -y + \frac{a p_y}{1 + b\left(p_x^2 + p_y^2\right)},
    \end{aligned}
\end{equation}
where \(a>0\) and \(b>0\) are parameters.
The map admits two independent invariants,
\begin{equation}
    \begin{gathered}
        \mathcal{K}_1
        =
        x^2 + y^2 + p_x^2 + p_y^2
        - a(xp_x + yp_y)
        + b(xp_x + yp_y)^2, \\
        \mathcal{K}_2
        =
        xp_y - yp_x .
    \end{gathered}
\end{equation}

For the parameters \(a=1.6\) and \(b=1.0\), starting from the initial condition
\begin{equation}
    \mathbf{z}_0=\left(3.0,0.5,1.0,0.5\right)^\mathrm{T}, 
\end{equation}
application of the procedure described above yields the frequencies
\begin{equation}
    \begin{aligned}
        \nu_1&=0.461066585378995,\\
        \nu_2&=0.224317222882003
    \end{aligned}
\end{equation}
The values are reported with full numerical precision.

The McMillan map is separable in polar coordinates \cite{PhysRevAccelBeams.23.054001}
\begin{equation}
\begin{gathered}
    x=r\cos\theta,\qquad y=r\sin\theta\\
    r=\sqrt{x^2+y^2},\qquad \Delta\theta=\arctan\left(\frac{xp_y-yp_x}{xp_x+yp_y}\right)
\end{gathered}
\end{equation}
where the radial coordinate $r$ and the incremental angular advance $\Delta\theta$
oscillate with a single fundamental frequency $\nu_1$,  
while the polar angle $\theta$ increases linearly on average with rate $\nu_2$.

For comparison and calibration, the frequencies $\nu_{1,2}$ are determined
numerically from time-series data obtained by tracking with the same initial
condition.
The NAFF algorithm \cite{laskar2003frequencymapanalysisquasiperiodic} is applied to
the radial time series $\{r_t\}_{t=0}^{N}$ to extract $\nu_1$.
The angular frequency $\nu_2$ is obtained independently from the mean angular
advance,
\begin{equation}
    \nu_2
    =
    \frac{1}{2\pi N}
    \sum_{t=0}^{N-1} \Delta\theta_t ,
\end{equation}
which corresponds (for $N \rightarrow \infty$) to the rotation number associated with the polar angle.
The resulting frequencies are then compared with those obtained from the
invariant-flow method.

\begin{figure}[!htbp]
    \centering
    \includegraphics[width=0.99\columnwidth]{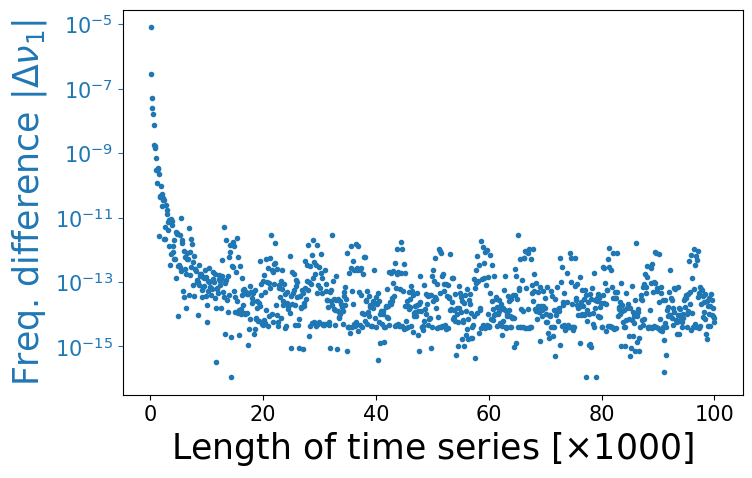}
    \includegraphics[width=0.99\columnwidth]{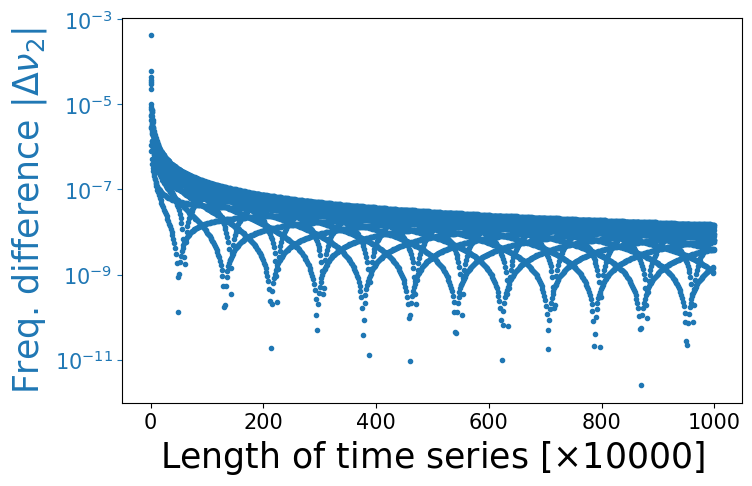}
    \caption{
        Comparison between time-series-based frequency extraction and the invariant-flow
        method for the McMillan map, top: $|\Delta\nu_1|$, bottom: $|\Delta\nu_2|$.
        As the tracking length increases, both numerical estimates converge to the
        invariant-flow results.
        Reducing the tracking length by one order of magnitude leads to a loss of three to
        four significant digits in $\nu_1$, while the estimate of $\nu_2$ converges more
        slowly and exhibits residual oscillatory deviations.
        The invariant-flow method achieves machine-level precision without relying on long
        time series.
    }
    \label{fig:naff_compare}
\end{figure}

Figure~\ref{fig:naff_compare} shows the magnitude of the frequency difference
$|\Delta \nu_{1,2}|$ as a function of the length of the time series.
As the time series length increases, both numerically extracted frequencies
$\nu_1$ and $\nu_2$ converge to the values obtained from the invariant-flow
method.
Reducing the time series length by one order of magnitude results in a loss of
three to four significant digits in the determination of $\nu_1$, while the
convergence of $\nu_2$ is significantly slower and exhibits residual oscillatory
deviations.
These oscillations decay gradually with increasing time-series length and arise
from bounded modulations superimposed on the mean angular advance.
In contrast, the invariant-flow method achieves high precision without relying
on long time series, since the frequencies are determined geometrically from
invariant flows rather than through spectral resolution.

The present approach is closely related to the framework developed by
C.~E.~Mitchell et al.~\cite{PhysRevE.103.062216}, in which fundamental frequencies are
obtained from integrals taken along arbitrary closed curves on the torus.
The invariant-flow formulation introduced here provides a constructive and
algorithmically explicit realization of this geometric principle.
By expressing closed curves as concatenations of commuting invariant flows,
the required integrals are replaced by time-of-flight parameters that are
determined directly through symplectic ordinary differential equation solvers
and optimization.
This leads to a simple and robust numerical procedure while retaining the
underlying geometric interpretation.
The present method fits naturally within the geometric framework of
Ref.~\cite{PhysRevE.103.062216}.

While the present formulation assumes a fully integrable system with exact
commuting invariants, many realistic systems are only near integrable, for which
approximate invariants can be constructed.
The invariant-flow construction can then be formally applied to these
approximate invariants to yield effective frequency estimates.
Because the corresponding flows no longer commute exactly, the extracted
frequencies are generally approximate.
The resulting deviation provides a quantitative signature of non-integrability
and depends on the quality of the approximate invariants. A systematic study of this regime is beyond the scope of the present work
and will be the subject of future investigation.

An additional advantage of the invariant-flow formulation is that it does not
require analytic expressions for the invariants themselves.
The method relies only on the evaluation of the Hamiltonian vector fields
generated by the invariants, $[\mathbf{z},\mathcal{K}_i]$, which depend on derivatives of
the invariants with respect to phase-space coordinates.
This makes the approach naturally compatible with data-driven or
neural-network-based representations of invariants, where gradients are readily
available through automatic differentiation.

In contrast to approaches that attempt to learn action-angle variables
directly, the present framework does not require enforcing global angle
periodicity or canonical structure during training.
Once approximate invariants are available, the frequencies can be extracted
directly from the corresponding invariant flows.
This suggests a natural route for combining modern data-driven techniques with
geometric frequency analysis.

In summary, the invariant-flow viewpoint clarifies the geometric origin of
frequencies in integrable systems by expressing them in terms of
time-of-flight parameters on phase-space tori.
This formulation provides a simple and robust numerical procedure rooted in the
geometry of invariant flows.


This work was supported by Brookhaven Science Associates, LLC, under Contract No. DE-SC0012704 with the U.S. Department of Energy; by a U.S. Department of Energy Early Career Award; by the DOE Basic Energy Sciences (BES) Field Work Proposal 2025-BNL-PS040; and by the DOE High Energy Physics (HEP) award DE-SC0019403.



\bibliography{ref}

\end{document}